\documentclass[iop, apj]{emulateapj}

\usepackage{natbib}
\usepackage{amsmath}
\usepackage{graphicx}

\usepackage[breaklinks, colorlinks, citecolor=blue, linkcolor=blue, urlcolor=blue]{hyperref}

\newcommand{\figuresize}{0.999}

\newcommand{\fig}[1]{Figure~\ref{fig:#1}}

\newcommand{\dex}{\,\textnormal{dex}}

\newcommand{\Msun}{\,\textnormal{M}_\odot}

\newcommand{\Mpc}{\,\textnormal{Mpc}}

\newcommand{\kpc}{\,\textnormal{kpc}}

\newcommand{\pc}{\,\textnormal{pc}}
\newcommand{\cm}{\,\textnormal{cm}}

\newcommand{\Gyr}{\,\textnormal{Gyr}}

\newcommand{\yri}{\,\textnormal{yr}^{-1}}
\newcommand{\kms}{\,\textnormal{km}\,\textnormal{s}^{-1}}

\newcommand{\K}{\,\textnormal{K}}

\newcommand{\matter}{\textnormal{matter}}

\newcommand{\dm}{\textnormal{dm}}
\newcommand{\baryon}{\textnormal{baryon}}
\newcommand{\gas}{\textnormal{gas}}

\newcommand{\stars}{\textnormal{star}}

\newcommand{\host}{\textnormal{host}}

\newcommand{\FeH}{\textnormal{[Fe/H]}}

\newcommand{\sfr}{\textnormal{SFR}}

\newcommand{\Mstar}{M_\textnormal{star}}

\newcommand{\Mhalo}{M_\textnormal{halo}}

\newcommand{\vcirc}{v_\textnormal{circ}}
\newcommand{\Vcircmax}{V_\textnormal{circ,max}}

\newcommand{\Mvir}{M_\textnormal{vir}}

\newcommand{\Mthm}{M_{200\textnormal{m}}}
\newcommand{\Rthm}{R_{200\textnormal{m}}}

\newcommand{\sigmavelstar}{\sigma_\textnormal{velocity,star}}

\begin{document}

\journalinfo{The Astrophysical Journal Letters, submitted}

\title{Reconciling dwarf galaxies with $\Lambda$CDM cosmology: \\
Simulating a realistic population of satellites around a Milky Way-mass galaxy}
\shorttitle{Reconciling dwarf galaxies with $\Lambda$CDM cosmology}
\shortauthors{Wetzel et al.}

\author{Andrew R. Wetzel\altaffilmark{1,2,3,8,9}}

\author{Philip F. Hopkins\altaffilmark{1}}

\author{Ji-hoon Kim\altaffilmark{1,4,10}}

\author{Claude-Andr{\'e} Faucher-Gigu{\`e}re\altaffilmark{5}}

\author{Du{\v s}an Kere{\v s}\altaffilmark{6}}

\author{Eliot Quataert\altaffilmark{7}}

\altaffiltext{1}{TAPIR, California Institute of Technology, Pasadena, CA, USA}
\altaffiltext{2}{Carnegie Observatories, Pasadena, CA, USA}
\altaffiltext{3}{Department of Physics, University of California, Davis, CA, USA}
\altaffiltext{4}{Kavli Institute for Particle Astrophysics and Cosmology, Department of Physics, Stanford University, Stanford, CA, USA}
\altaffiltext{5}{Department of Physics and Astronomy and CIERA, Northwestern University, Evanston, IL, USA}
\altaffiltext{6}{Department of Physics, Center for Astrophysics and Space Sciences, University of California, San Diego, CA, USA}
\altaffiltext{7}{Department of Astronomy and Theoretical Astrophysics Center, University of California, Berkeley, CA, USA}
\altaffiltext{8}{Moore Prize Fellow}
\altaffiltext{9}{Carnegie Fellow in Theoretical Astrophysics}
\altaffiltext{10}{Einstein Fellow}

\begin{abstract}
Low-mass ``dwarf'' galaxies represent the most significant challenges to the cold dark matter (CDM) model of cosmological structure formation.
Because these faint galaxies are (best) observed within the Local Group (LG) of the Milky Way (MW) and Andromeda (M31), understanding their formation in such an environment is critical.
We present first results from the Latte Project: the Milky Way on FIRE (Feedback in Realistic Environments).
This simulation models the formation of a MW-mass galaxy to $z \! = \! 0$ within $\Lambda$CDM cosmology, including dark matter, gas, and stars at unprecedented resolution: baryon particle mass of $7070 \Msun$ with gas kernel/softening that adapts down to $1 \pc$ (with a median of $25 \! - \! 60 \pc$ at $z \! = \! 0$).
Latte was simulated using the \textsc{GIZMO} code with a mesh-free method for accurate hydrodynamics and the FIRE-2 model for star formation and explicit feedback within a multi-phase interstellar medium.
For the first time, Latte self-consistently resolves the spatial scales corresponding to half-light radii of dwarf galaxies that form around a MW-mass host down to $\Mstar \! \gtrsim \! 10 ^ 5 \Msun$.
Latte's population of dwarf galaxies agrees with the LG across a broad range of properties: (1) distributions of stellar masses and stellar velocity dispersions (dynamical masses), including their joint relation; (2) the mass-metallicity relation; and (3) a diverse range of star-formation histories, including their mass dependence.
Thus, Latte produces a realistic population of dwarf galaxies at $\Mstar \! \gtrsim \! 10 ^ 5 \Msun$ that does \textit{not} suffer from the ``missing satellites'' or ``too big to fail'' problems of small-scale structure formation.
We conclude that baryonic physics can reconcile observed dwarf galaxies with standard $\Lambda$CDM cosmology.
\end{abstract}


\section{Introduction}

Dwarf galaxies ($\Mstar \! \lesssim \! 10 ^ 9 \Msun$) provide the smallest-scale probes of cosmological structure formation and thus are compelling laboratories to test the cold dark matter (CDM) framework.
However, observed dwarf galaxies in the Local Group (LG) of the Milky Way (MW) and Andromeda (M31) present significant challenges to CDM.
First, the ``missing satellites'' problem: far fewer luminous satellites appear to be observed around the MW than dark-matter-only models predict \citep{Moore1999, Klypin1999b}.
More concretely, the ``too big to fail'' problem: dark-matter-only simulations predict too many massive dense subhalos compared with satellites around the MW \citep{Read2006, BoylanKolchin2011}. 
Relatedly, the ``core-cusp'' problem: the inner density profiles of dwarf galaxies appear to be cored, rather than cuspy as CDM predicts \citep[e.g.,][]{FloresPrimack1994, Moore1994, Simon2005, Oh2011}.

Many works have explored modifications to standard CDM, such as warm \citep[e.g.,][]{Lovell2014} or self-interacting \citep[e.g.,][]{Rocha2013} dark matter.
However, one must account for baryonic physics as well, and many theoretical studies have shown that stellar feedback can drive strong gas inflows/outflows that generate significant dark-matter cores
in dwarf galaxies \citep[e.g.,][]{ReadGilmore2005, Mashchenko2008, PontzenGovernato2012, DiCintio2014, Chan2015}.
Almost all baryonic simulations have modeled isolated dwarf galaxies, which are computationally tractable at the necessary resolution.
However, faint dwarf galaxies are (most robustly) observed near the MW/M31.
Thus, modeling dwarf-galaxy formation within such a host-halo environment is critical, both to understand the role of this environment in their formation and to provide the proper environment to compare statistical properties of the population against the LG.

Cosmological simulations of MW-mass galaxies have progressed at increasing resolution and with more realistic stellar physics \citep[e.g.,][]{Hopkins2014a, AgertzKravtsov2015, Mollitor2015}, and some simulations have started to resolve the more massive satellite galaxies within MW-mass halos, with promising results \citep[e.g.,][]{BrooksZolotov2014, Sawala2016}.
However, such simulations have not yet achieved sufficiently high spatial resolution (comparable to simulations of isolated dwarf galaxies) to robustly resolve the half-light radii of such satellites, as small as $\sim \! 200 \pc$.

In this Letter, we introduce the Latte Project: the Milky Way on FIRE (Feedback in Realistic Environments).
Our goal is to simulate a series of MW-mass galaxies to $z \! = \! 0$ within $\Lambda$CDM cosmology at sufficient resolution to resolve both the host galaxy and dwarf galaxies that form around it, including state-of-the-art mesh-free hydrodynamics and the FIRE-2 model for stellar physics.
Here, we present first results, focusing on the dwarf-galaxy population.
In subsequent papers, we will examine more detailed properties of dwarf galaxies, including dark-matter profiles and gas content.

\section{Latte Simulations}

\subsection{GIZMO code and FIRE-2 model}

We run our simulations using the code \textsc{GIZMO}\footnote{http://www.tapir.caltech.edu/${\sim}$phopkins/Site/GIZMO} \citep{Hopkins2015} with the FIRE model for star formation and explicit feedback \citep{Hopkins2014a}.
In this Letter, we introduce several \textit{numerical} improvements to FIRE, detailed in Hopkins et al., in prep.
We refer to this improved implementation as ``FIRE-2''.

\textsc{GIZMO} uses a TREE+PM gravity solver updated from \textsc{GADGET}-3 \citep{Springel2005e}.
For hydrodynamics, we now use the mesh-free finite-mass (MFM) method, which is Lagrangian and provides automatically adaptive spatial resolution while maintaining machine-level conservation of mass, energy, and momentum, and excellent conservation of angular momentum.
MFM simultaneously captures advantages of both Lagrangian smooth-particle hydrodynamics (SPH) and Eulerian adaptive mesh refinement (AMR) schemes \citep[see][]{Hopkins2015}.

We incorporate radiative cooling and heating rates from \textsc{CLOUDY} \citep{Ferland2013} across $10 \! - \! 10 ^ {10} \K$, including atomic, molecular, and metal-line cooling for 11 elements.
We include ionization/heating from a redshift-dependent, spatially uniform ultraviolet background, including cosmic reionization, from \citet{FaucherGiguere2009}.

Stars form only in locally self-gravitating, molecular gas \citep{Hopkins2013c} at densities of $n_{\rm SF} \! > \! 1000 \cm ^ {-3}$ with instantaneous efficiency of 100\% per free-fall time, and the maximum density reached is $\approx \! 10 ^ 7 \cm^{-3}$ (corresponding to $h_\gas \! \approx \! 1 \pc$).
We incorporate a comprehensive set of stellar feedback processes: radiation pressure from massive stars, local photoionization and photoelectric heating, stellar winds, core-collapse and Ia supernovae.
We compute values for mass, momentum, and thermal energy injection \textit{directly} from \textsc{STARBURST99} v7.0 \citep{Leitherer1999}.

We note the most important improvements in FIRE-2 as compared with previous FIRE simulations.
First, FIRE-2 uses MFM instead of pressure-entropy smoothed particle hydrodynamics (P-SPH), because MFM offers superior performance across a range of tests \citep{Hopkins2015}.
That said, several tests (\citealt{Hopkins2015, Dave2016}; Hopkins et al., in prep.) also show that using MFM versus P-SPH does not significantly change ($\lesssim \! 20\%$) the stellar properties of dwarf galaxies.
Second, FIRE-2 uses $n_{\rm SF} \! > \! 1000 \cm ^ {-3}$, higher than $n_{\rm SF} \! > \! 100 \, h^2 \, \cm^{-3} \! = \! 50 \cm ^ {-3}$ used previously.
However, we tested using $n_{\rm SF} \! > \! 5 \! - \! 1000 \cm ^ {-3}$ and found no significant differences in galaxy properties, because our most important criterion is that star-forming gas be \textit{locally self-gravitating} (so even for $n_{\rm SF} \! > \! 50 \cm ^ {-3}$, the average density of star-forming gas is $\sim \! 1000 \cm ^ {-3}$).
Finally, FIRE-2 improves how FIRE numerically \textit{couples} stellar feedback to surrounding gas.
Previously, FIRE coupled a star particle's feedback into the nearest $\approx \! 32$ gas particles, each receiving a fraction \textit{regardless} of their geometric distribution around the star particle, which could lead to non-conservation of (net) momentum.
FIRE-2 couples feedback to all gas particles whose kernel encompasses the star particle, with a fraction proportional to the subtended solid angle, and renormalizes each gas particle's fraction along each spatial dimension, ensuring that the total injection of mass, energy, and (net) momentum are conserved to machine accuracy.
In Hopkins et al., in prep., we describe these improvements in detail, showing that they can affect the stellar morphology of MW-mass galaxies, but they have no significant effect on dwarf galaxies, which are the focus of this Letter.

Previous FIRE cosmological simulations of \textit{isolated} dwarf galaxies have reproduced several key observables: realistic galactic outflows \citep{Muratov2015}, $\Mstar$-metallicity relation \citep{Ma2016}, $\Mstar$-size relation \citep{ElBadry2016}, cored dark-matter profiles \citep{Onorbe2015, Chan2015}, and dispersion-dominated stellar kinematics \citep{Wheeler2015b}.

\subsection{Cosmological Zoom-in Simulations}

We cosmologically simulate a MW-mass halo at high resolution using the zoom-in technique \citep[see][]{Onorbe2014}.
We first run a dark-matter-only simulation within a periodic volume of length $85.5 \Mpc$ with $\Lambda$CDM cosmology: $\Omega_\Lambda \! = \! 0.728$, $\Omega_\matter \! = \! 0.272$, $\Omega_\baryon \! = \! 0.0455$, $h \! = \! 0.702$, $\sigma_8 \! = \! 0.807$, and $n_s \! = \! 0.961$.
We select at $z \! = \! 0$ an isolated halo with $\Rthm \! = \! 334\kpc$ (within which the mass density is $200 \times$ the average matter density), $\Mthm \! = \! 1.3 \times 10 ^ {12} \Msun$ ($\Mvir \! = \! 1.1 \times 10 ^ {12} \Msun$), and maximum circular velocity $\Vcircmax \! = \! 162 \kms$.
This is the same ``m12i'' halo from \citet{Hopkins2014a}.
We trace particles within $5 \, \Rthm$ back to $z \! = \! 100$ and regenerate the encompassing convex hull at high resolution, embedded within the full lower-resolution volume, using the \textsc{MUSIC} code \citep{HahnAbel2011}.
Rerun to $z \! = \! 0$, this zoom-in region has zero low-resolution contamination within $d_\host < 600 \kpc$ ($1.8 \, \Rthm$).

Our fiducial baryonic simulation contains dark matter, gas, and stars within the zoom-in region, comprising 140 million total particles, with $m_\dm \! = \! 3.5 \times 10 ^ 4 \Msun$ and $m_\textnormal{gas,initial} \! = \! m_\textnormal{star,initial} \! = \! 7070 \Msun$.
Dark matter and stars have fixed gravitational softening: $h_\dm \! = \! 20 \pc$ and $h_\stars \! = \! 4 \pc$ (Plummer equivalent), comoving at $z \! > \! 9$ and physical thereafter.
Gas smoothing is fully adaptive \citep[see][]{Hopkins2015} and is the same for the hydrodynamic kernel and gravitational softening.
The smallest gas kernel/softening achieved is $h_\gas \! = \! 1.0 \pc$; the median within the host galaxy and (gaseous) dwarf galaxies is $\approx \! 25$ and $\approx \! 60 \pc$, respectively, at $z \! = \! 0$.
These softenings allow us to measure properties like velocity dispersions and dynamical masses at our dwarf galaxies' half-light radii ($r \! \gtrsim \! 200 \pc$) with good resolution ($> \! 10 \, h_\dm, > \! 40 \, h_\stars$).\footnote{
We find that $\Mstar$ of dwarf galaxies converges to within $0.3 \dex$ above $\approx \! 8$ star particles (Hopkins et al., in prep.), and total density converges to $\approx \! 0.2 \dex$ at the radius enclosing $\sim \! 200$ total particles, which is satisfied within the half-light radii of the dwarf galaxies here.
See also \citet{Onorbe2015, Chan2015}.}
Particle time-stepping is fully adaptive: the shortest time-step achieved is 180 years.

We ran this simulation on the Stampede supercomputer using 2048 cores for 15 days (720,000 CPU-hours).
To test the effects of baryonic physics and numerical resolution, we also ran a (1) dark-matter-only simulation at the same resolution and (2) baryonic simulation with $8 \times$ larger particle mass ($m_\dm \! = \! 2.8 \times 10 ^ 5 \Msun$, $m_\textnormal{gas,initial} \! = \! 5.7 \times 10 ^ 4 \Msun$) and $2 \times$ larger force softenings.

To identify (sub)halos and their galaxies, we use a modified version of the six-dimensional phase-space halo finder \textsc{rockstar}\footnote{https://bitbucket.org/pbehroozi/rockstar-galaxies} \citep{Behroozi2013a}, which accounts for multiple species and assigns dark-matter, gas, and star particles to (sub)halos.

\section{Results}

\renewcommand{\figuresize}{0.363}
\begin{figure*}
\centering
\includegraphics[height = \figuresize \textwidth]{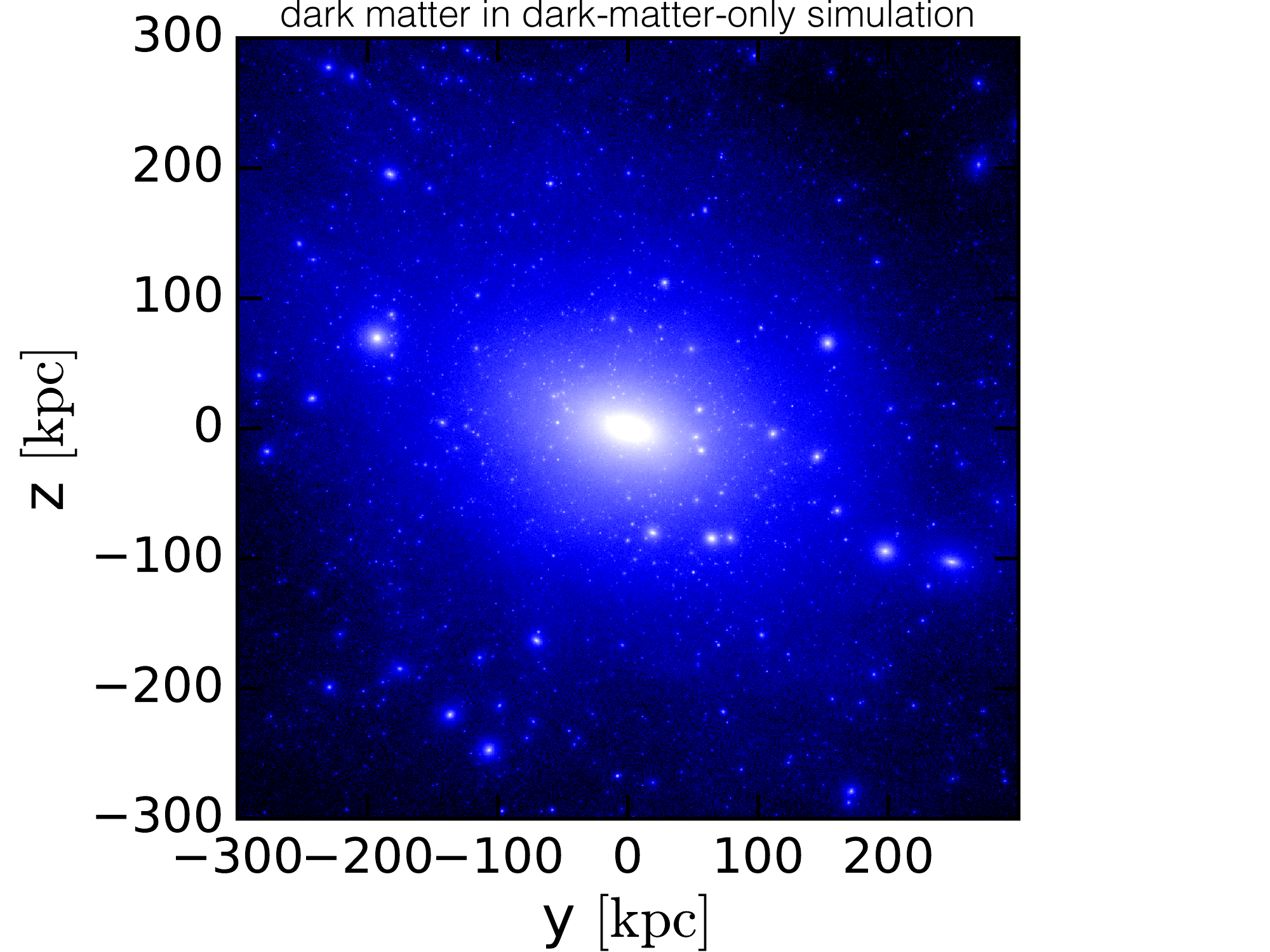}
\hspace{-2 mm}
\includegraphics[height = \figuresize \textwidth]{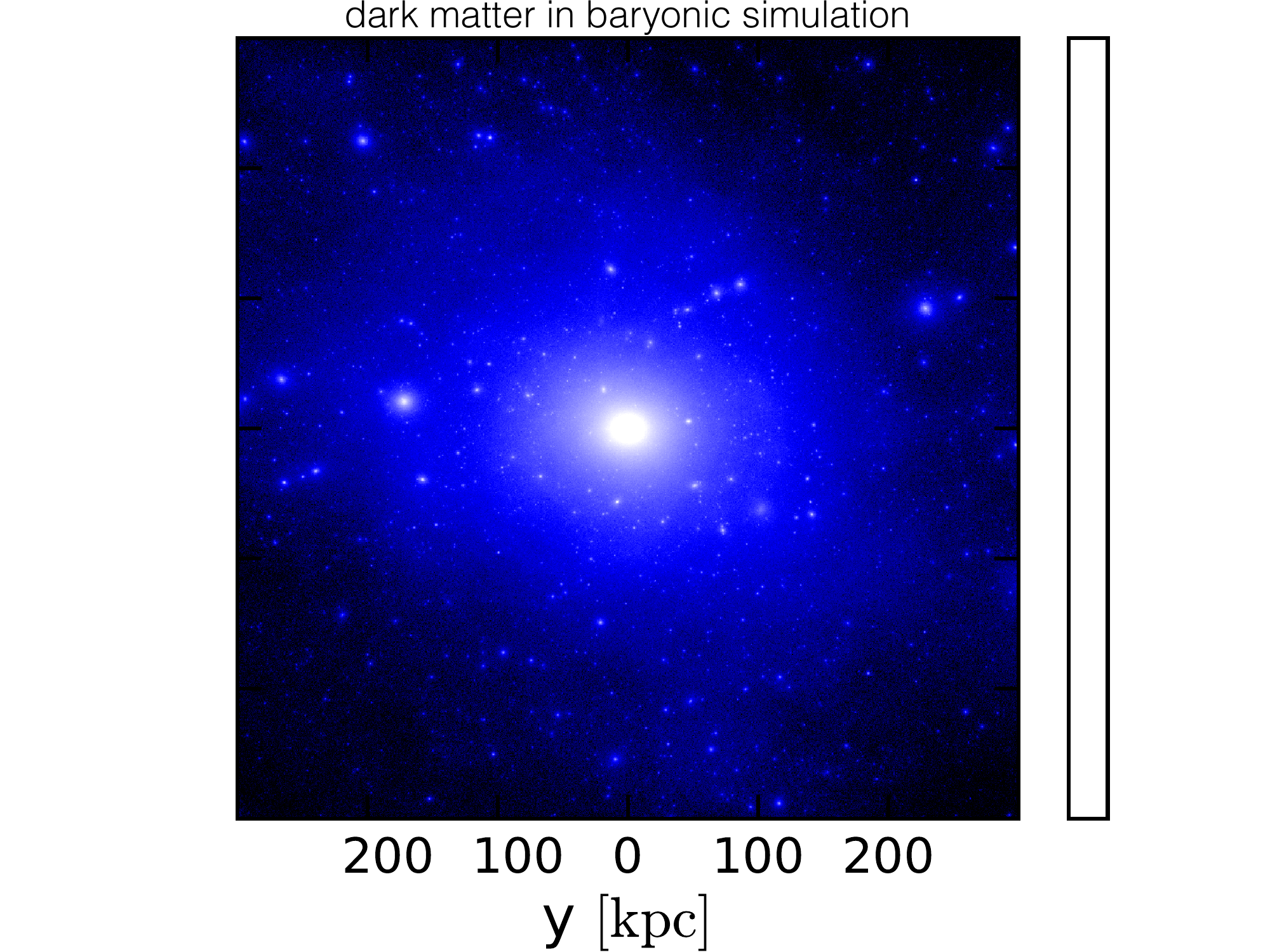}
\hspace{-2 mm}
\includegraphics[height = \figuresize \textwidth]{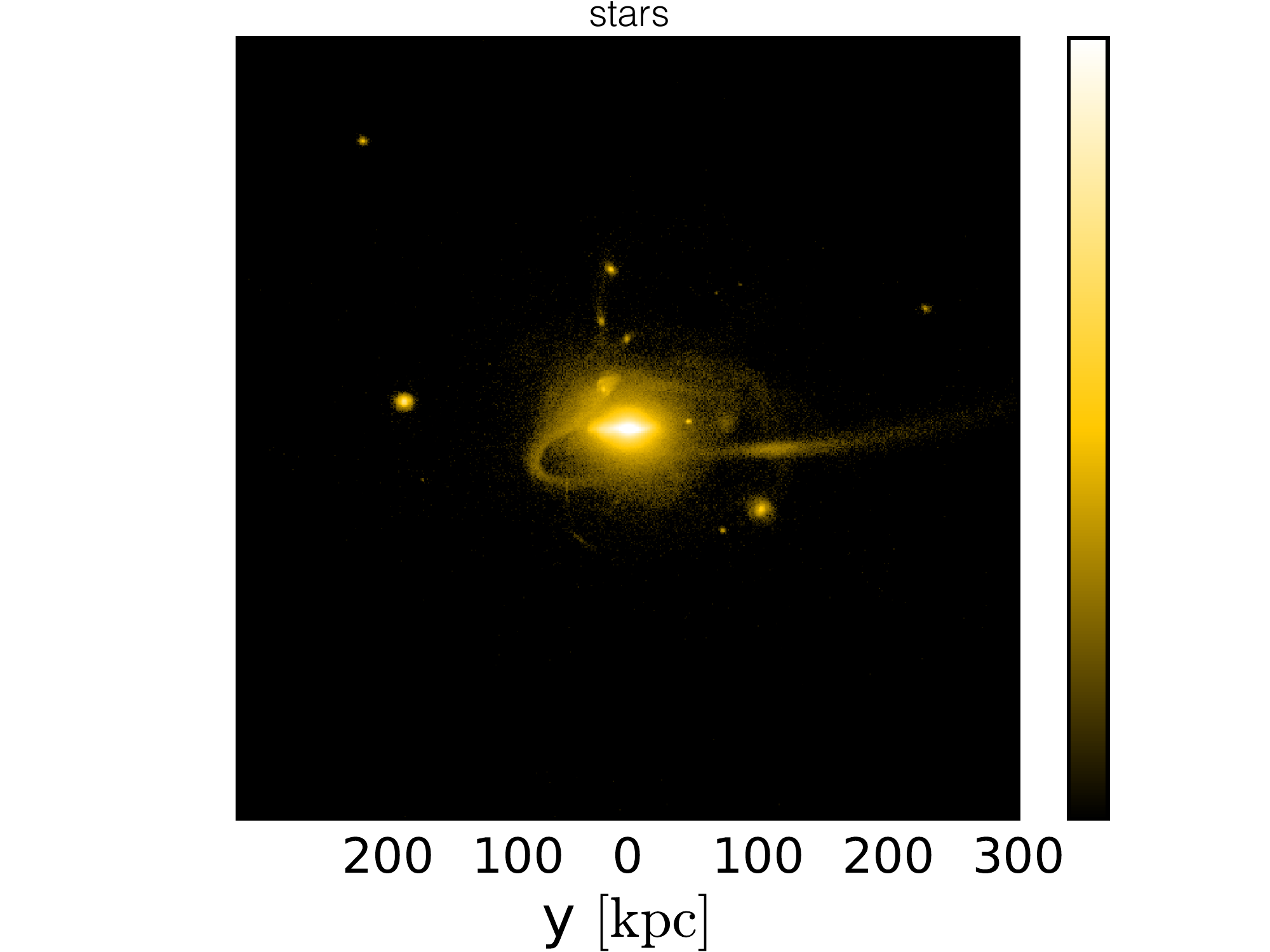}
\caption{
Projected surface densities around the MW-mass host in the Latte simulation at $z \! = \! 0$: the dark-matter-only simulation (left); dark matter (middle) and stars (right) in the baryonic simulation.
Color scales are logarithmic, both spanning $10 ^ 4 \! - \! 10 ^ 8 \Msun \kpc ^ {-2}$.
The baryonic simulation contains $\approx \! 3 \times$ fewer subhalos than the dark-matter-only simulation at fixed $\Vcircmax$, with 13 satellite galaxies at $\Mstar \! > \! 8 \times 10 ^ 4 \Msun$.
}
\label{fig:image}
\end{figure*}

\renewcommand{\figuresize}{0.398}
\begin{figure*}
\centering
\includegraphics[height = \figuresize \textwidth]{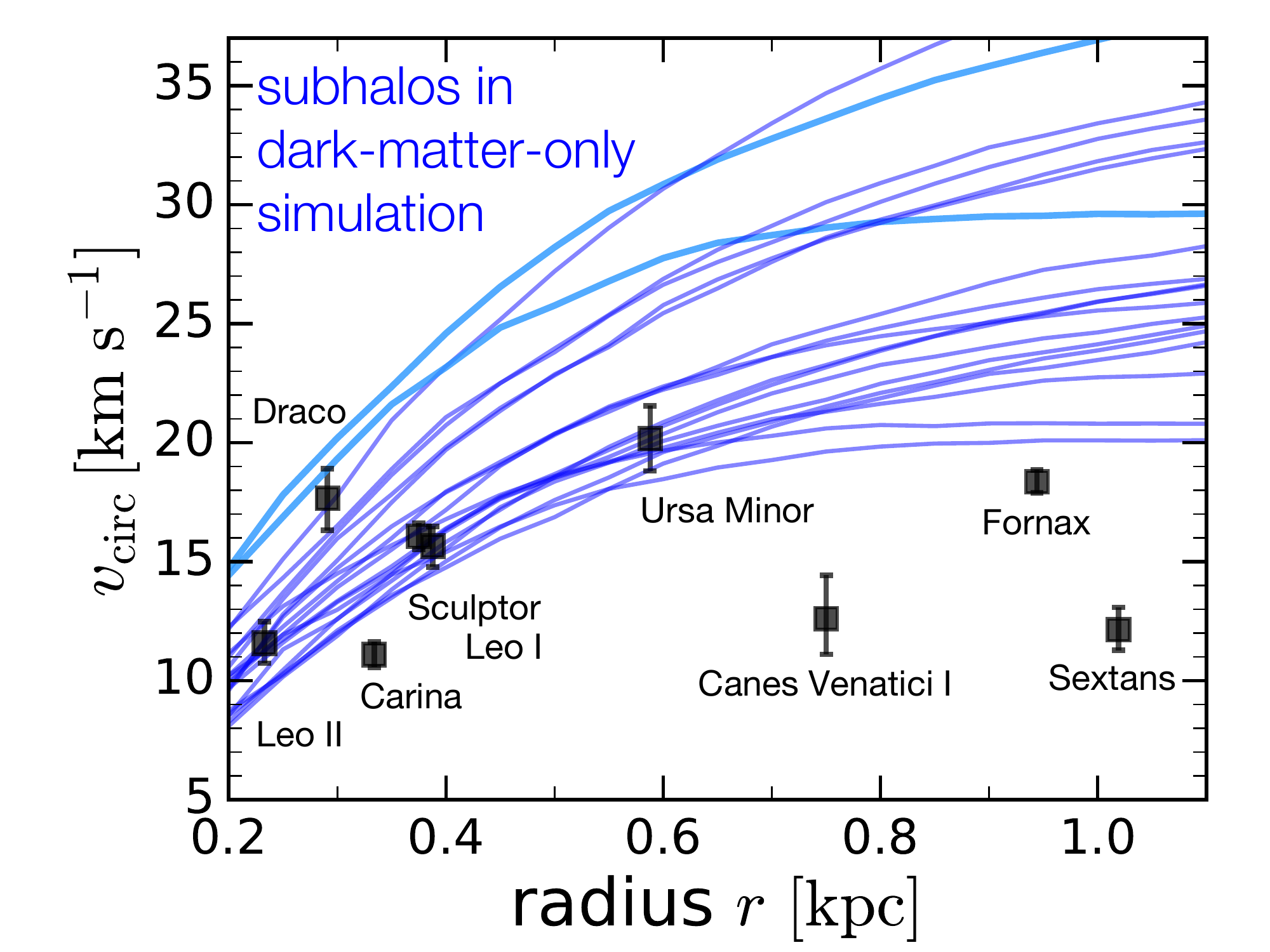}
\includegraphics[height = \figuresize \textwidth]{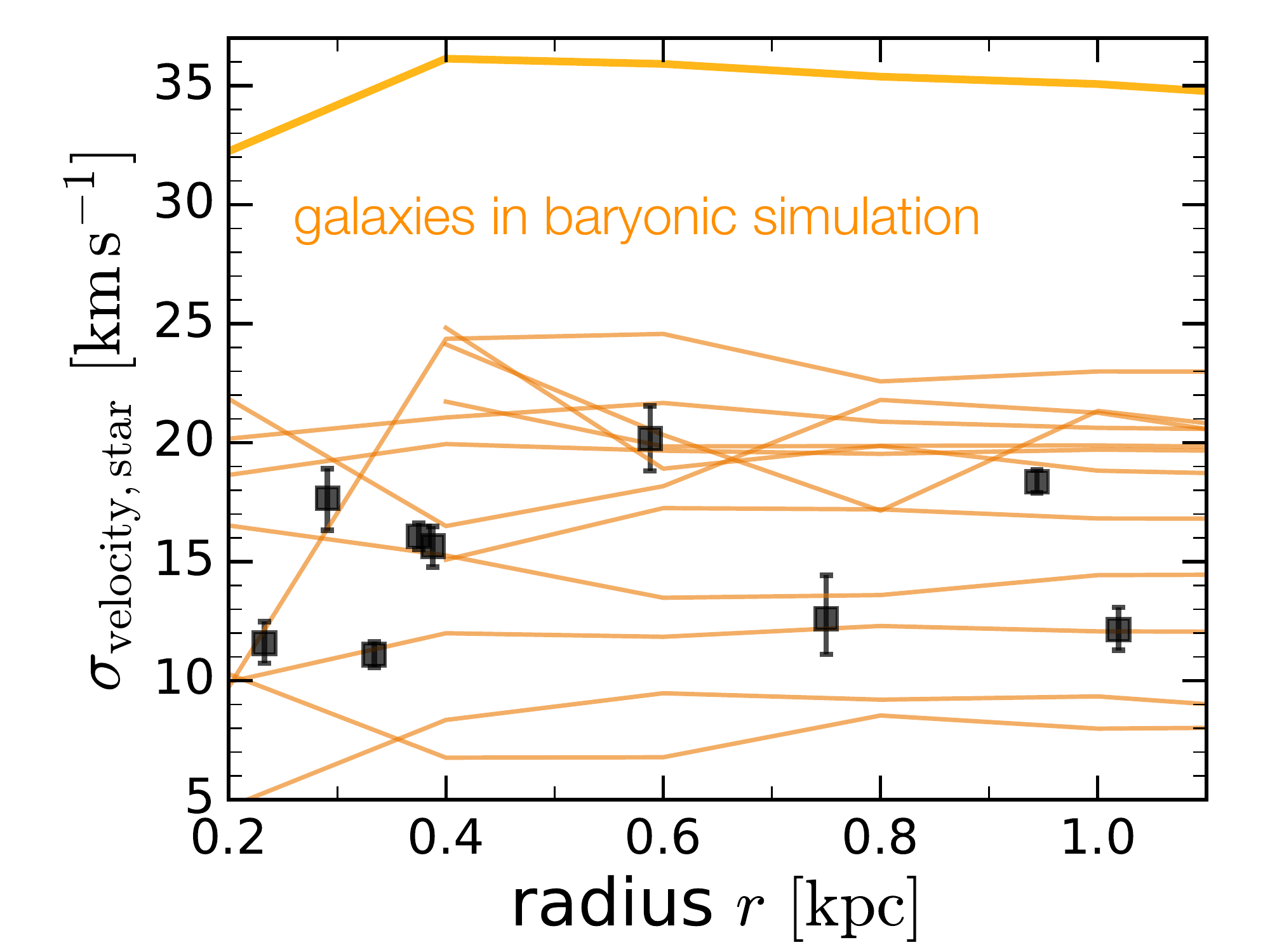}
\caption{
\textbf{Left}:
profiles of circular velocity, $\vcirc(r) \! = \! \sqrt{G m_\textnormal{total}(< r) / r}$ at $z \! = \! 0$. 
Points show observed satellites of the MW with $\Mstar \! = \! 2 \times 10 ^ 5 \! - \! 2 \times 10 ^ 7 \Msun$ \citep{Wolf2010}.
Curves show the 19 subhalos in the dark-matter-only simulation at $d_\host \! < \! 300 \kpc$ with densities as low as Ursa Minor.
Two subhalos (light blue) are denser than all observed satellites.
Allowing one to host the SMC and noting that 5 others are consistent with Ursa Minor, Draco, Sculptor, Leo~I, and Leo~II leads to 13 that are too dense (the ``too big to fail'' problem).
\textbf{Right}:
profiles of stellar 3D velocity dispersion for the 13 satellite galaxies in the baryonic simulation.
All profiles are nearly flat with radius.
One satellite has high dispersion, closer to the SMC's $48 \kms$; all others are broadly consistent with the MW.
}
\label{fig:vel.circ_v_radius}
\end{figure*}

\renewcommand{\figuresize}{0.38}
\begin{figure*}
\centering
\includegraphics[height = \figuresize \textwidth]{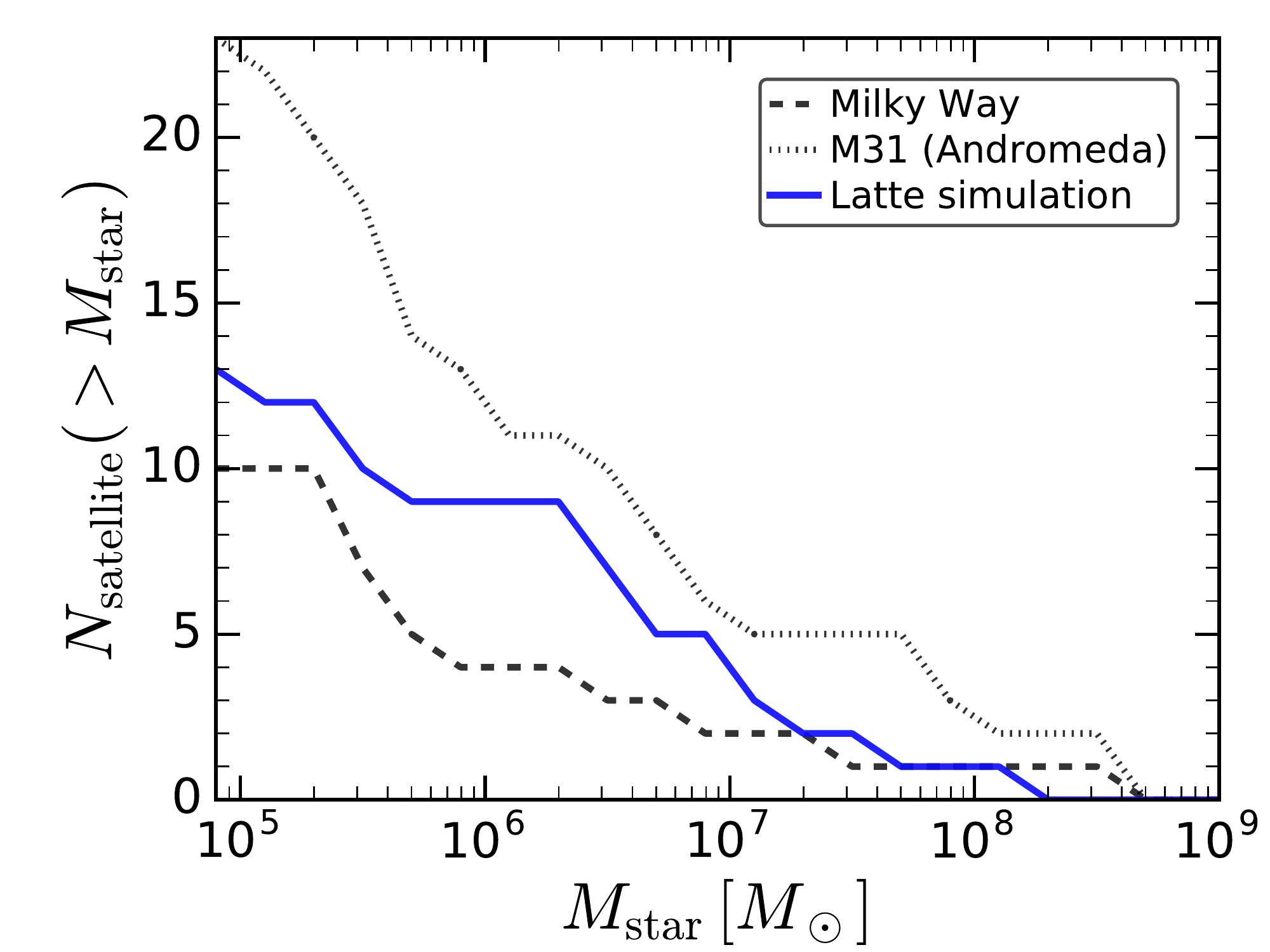}
\includegraphics[height = \figuresize \textwidth]{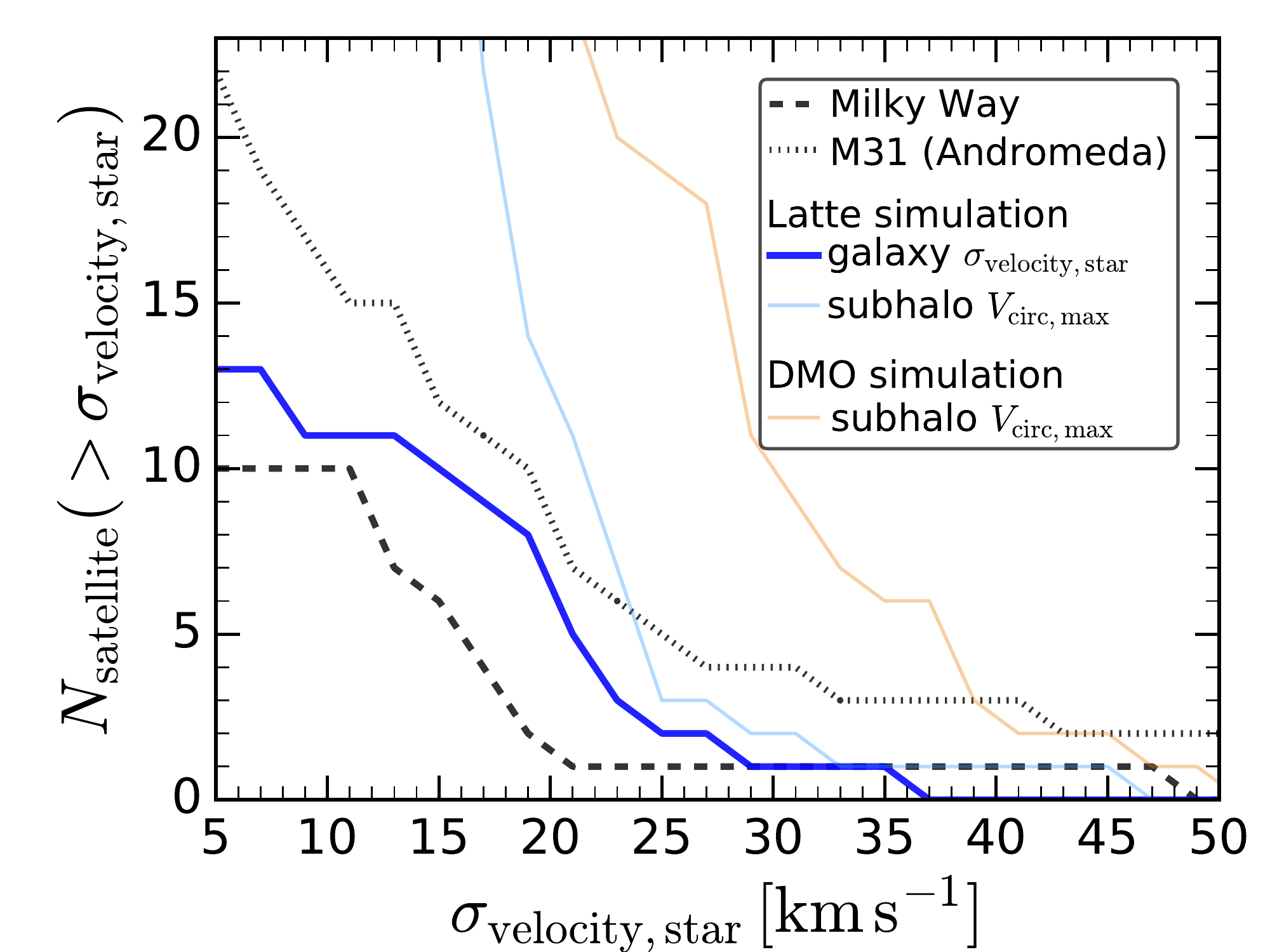}
\caption{
Cumulative number of satellites at $z \! = \! 0$ above a given stellar mass (left) and stellar 3D velocity dispersion (right) in the Latte simulation (blue) and observed around the Milky Way (MW; dashed) and Andromeda (M31; dotted), excluding the LMC, M33, and Sagittarius.
For both $\Mstar$ and $\sigma$, Latte's satellites lie entirely between the MW and M31, so \textit{Latte does not suffer from the ``missing satellites'' or ``too big to fail'' problems}.
Thin curves (right) show $\Vcircmax$ for \textit{all} dark-matter subhalos in the baryonic (light blue) and dark-matter-only (DMO; orange) simulations, demonstrating the $\approx \! 3 \times$ reduction from baryonic physics.
}
\label{fig:mass_function}
\end{figure*}

\renewcommand{\figuresize}{0.999}
\begin{figure}
\centering
\includegraphics[width = \figuresize \columnwidth]{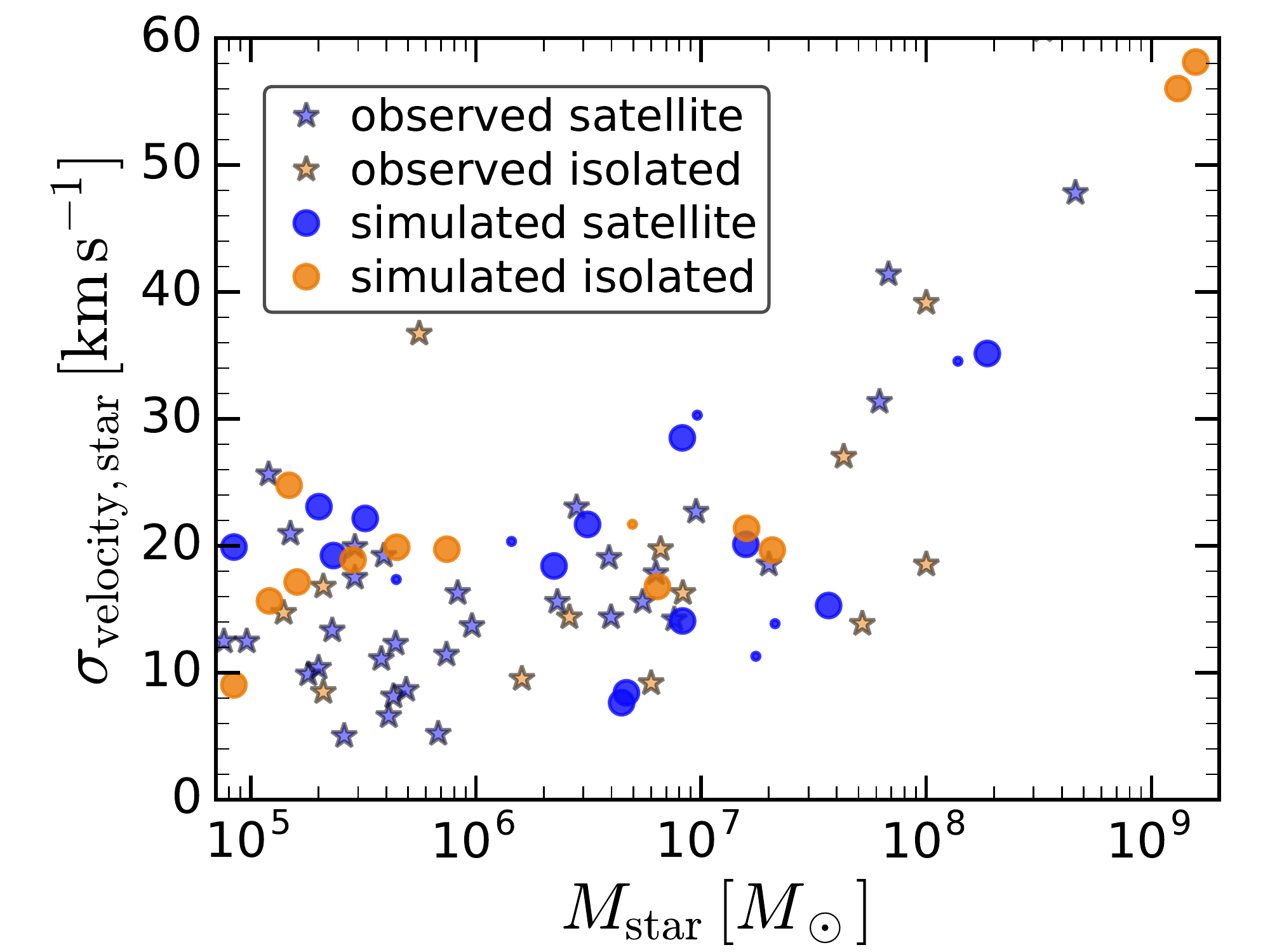}
\includegraphics[width = \figuresize \columnwidth]{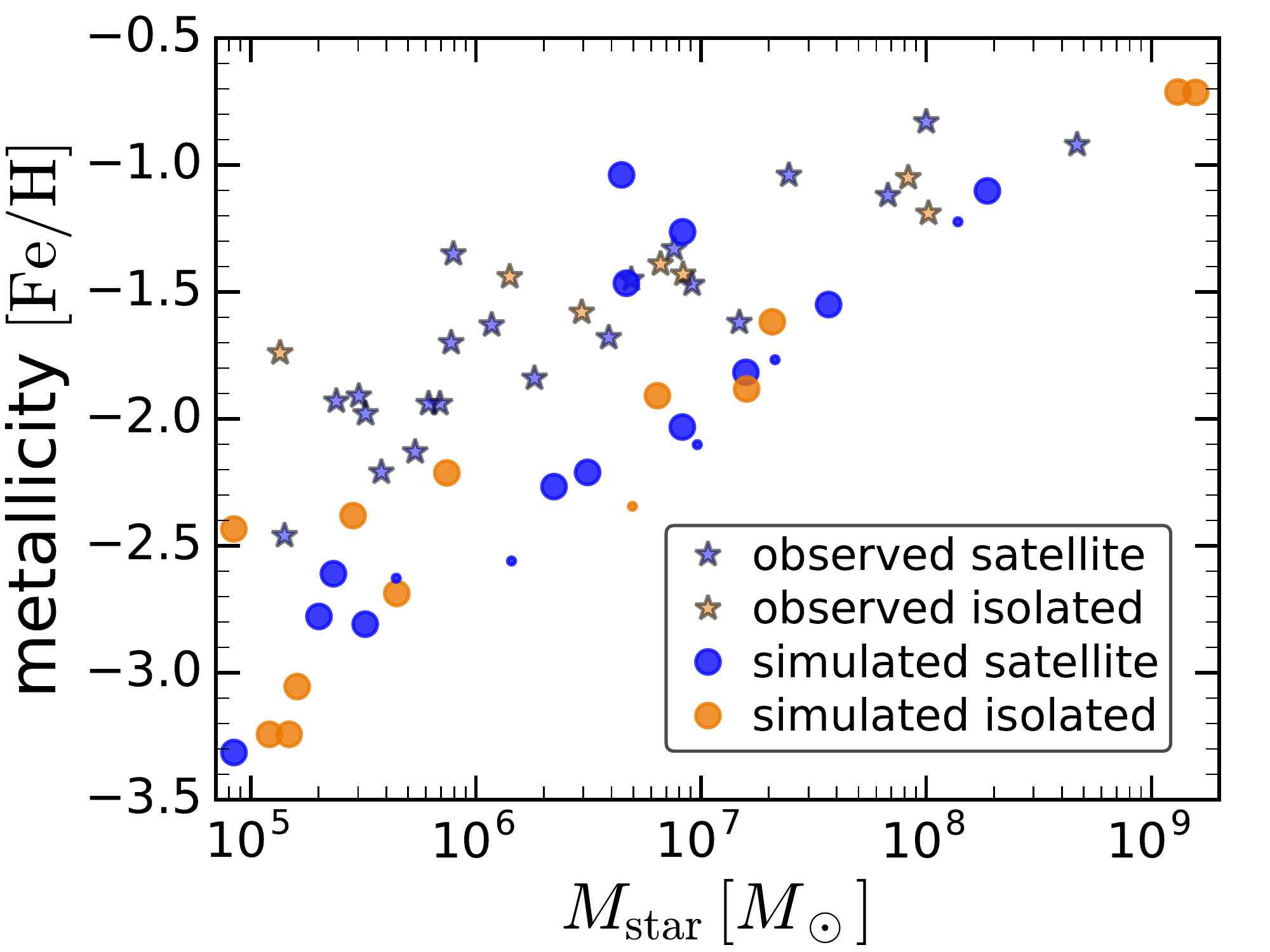}
\caption{
Properties of dwarf galaxies versus stellar mass, for both satellite ($d_\host \! < \! 300 \kpc$, blue) and isolated ($d_\host \! > \! 300 \kpc$, orange) galaxies in Latte (circles) and observed around the MW and M31 (stars).
Small circles show the lower-resolution simulation.
\textbf{Top}: stellar 3D velocity dispersion.
Latte's galaxies agree with observations in their \textit{joint} relation between $\Mstar$ and $\sigmavelstar$.
The lower-resolution simulation agrees well down to $\Mstar \! \approx \! 6 \times 10 ^ 5 \Msun$.
\textbf{Bottom}: stellar iron metallicity, scaled to solar, $\FeH$, compared with observations \citep{Kirby2013b}.
Latte's galaxies show a clear $\Mstar$-metallicity relation, with no significant offset between satellite and isolated galaxies (except 3 satellites), as observed.
$\FeH$ broadly agrees with observations at $\Mstar \! \gtrsim \! 5 \times 10 ^ 6 \Msun$ though is $\approx \! 0.5 \dex$ low at lower $\Mstar$.
}
\label{fig:property_v_mass}
\end{figure}

\renewcommand{\figuresize}{0.999}
\begin{figure}
\centering
\includegraphics[width = \figuresize \columnwidth]{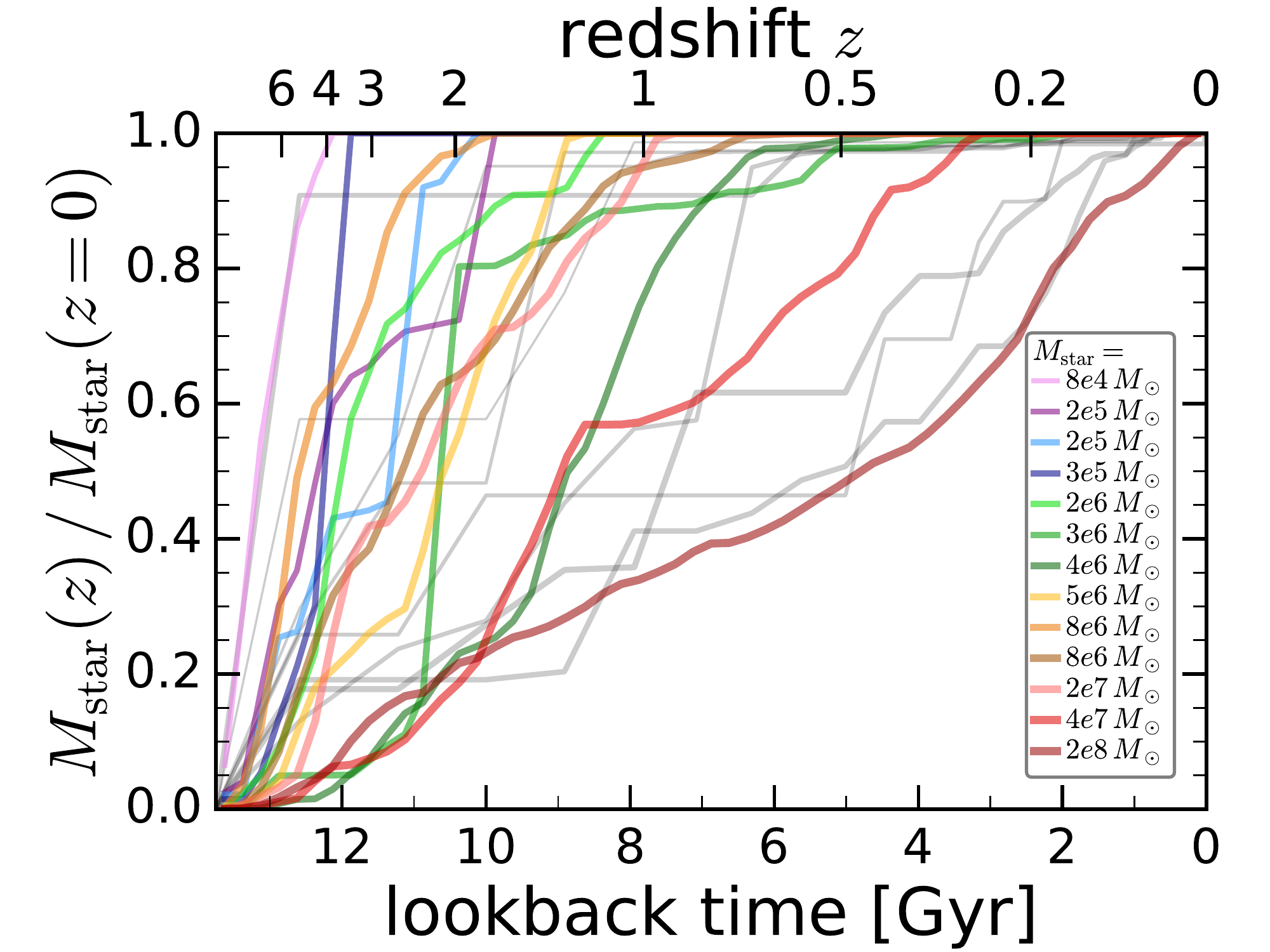}
\caption{
Cumulative star-formation histories, $\Mstar(z) / \Mstar(z \! = \! 0)$, of satellites at $z \! = \! 0$.
Colored curves show the Latte simulation, with $\Mstar(z \! = \! 0)$ labeled.
Gray curves show observed satellites of the MW (same as \fig{vel.circ_v_radius}, excepting Sextans) from \citet{Weisz2014a}, with line thickness indicating $\Mstar(z \! = \! 0)$ across $2 \times 10 ^ 5 \! - \! 2 \times 10 ^ 7 \Msun$.
Latte's satellites experienced a broad range of star-formation histories, and all have quenched (turned off) star formation by $z\! \approx \! 0$, except the most massive.
Both trends are consistent with observations.
}
\label{fig:star_formation_history}
\end{figure}

At $z \! = \! 0$ the MW-mass host galaxy has $\Mstar \! = \! 7 \times 10 ^ {10} \Msun$, with a bulge-to-disk mass ratio of 1:7.
The star-formation rate $\sfr \! = \!6 \Msun \yri$, which gradually declined from a peak $20 \Msun\yri$ at $z \! \approx \! 0.8$, leading to a late-forming galaxy (half of $\Mstar$ formed at $z \! < \! 0.6$).
For comparison, the MW has $\Mstar \! = \! 6 \times 10 ^ {10} \Msun$ and $\sfr \! = \! 1.7 \Msun \yri$ \citep{LicquiaNewman2015}; M31 has $\Mstar \! \approx \! 10 ^ {11} \Msun$ and $\sfr \! \approx \! 0.7 \Msun \yri$ \citep{Tamm2012, Lewis2015}.
Thus, the host's $\Mstar$ is close to the MW, though with $3.5 \times$ higher SFR.
We will examine the host in detail in Wetzel et~al., in prep.

At $z \! = \! 0$, \textsc{rockstar} identifies 25 dwarf galaxies down to $\Mstar \! > \! 8 \times 10 ^ 4 \Msun$ ($16$ star particles) within uncontaminated (sub)halos out to $d_\host \! = \! 3 \Mpc$.
We define ``satellite'' and ``isolated'' galaxies via $d_\host \! < \! 300 \kpc$ and $> \! 300 \kpc$, leading to 13 satellite and 12 isolated dwarf galaxies.
For the latter, the minimum halo mass is $\Mthm \! \approx \! 10 ^ 9 \Msun$, so our dwarfs' halos are well resolved with $\gtrsim \! 40,000$ dark-matter particles (see \citealt{GarrisonKimmel2016} for their $\Mstar \! - \! \Mhalo$ relation).

\fig{image} shows the MW-mass halo at $z \! = \! 0$, showing surface densities $> \! 10 ^ 4 \Msun \kpc ^ {-2}$.
The dark-matter-only simulation (left) contains significant substructure.
By comparison, dark-matter substructure in the baryonic simulation (middle) is dramatically reduced at $d_\host \! < \! 300 \kpc$, where it contains $\approx \! 3 \times$ fewer subhalos at fixed $\Vcircmax$.
Furthermore, \fig{image} (right) shows that, of this reduced subhalo population, only 13 host a galaxy.
\fig{image} also highlights the significant stellar halo, including streams and shells from disrupting satellites.

To put baryonic physics in context, we first examine our dark-matter-only simulation.
\fig{vel.circ_v_radius} (left) shows profiles of $\vcirc(r) \! = \! \sqrt{G m_\textnormal{total}(<  r) / r}$ for satellites within $d_\host \! < \! 300 \kpc$.
We compare with observed dwarf galaxies, compiled in \citet{McConnachie2012}, limiting to $\Mstar \! > \! 10 ^ 5 \Msun$, where observational completeness is well understood \citep[e.g., Figure~1 in][]{Wetzel2015b}.
However, we exclude the LMC and M33, because such massive satellites are rare around a MW/M31-mass host \citep{Busha2011b, Tollerud2011}, and we exclude Sagittarius because it is disrupting into a stream.
Points show MW satellites ($\Mstar \! = \! 2 \times 10 ^ 5 \! - \! 2 \times 10 ^ 7 \Msun$) with well-measured dynamical masses from \citet{Wolf2010}.
Following \citet{GarrisonKimmel2014b}, we show $\vcirc(r)$ curves for the 19 subhalos that are at least as dense as Ursa Minor; these span $\Vcircmax \! = \! 20 \! - \! 51 \kms$.
Two subhalos (light blue) are denser than all observed satellites.
Allowing the highest $\Vcircmax$ subhalo to host the SMC, this leads to one ``failure'' (that is, denser than all observed satellites).
Furthermore, counting all other subhalos and subtracting the 5 that are consistent with Ursa Minor, Draco, Sculptor, Leo~I, and Leo~II, we find 13 subhalos that are denser than the MW's, consistent with the range measured in suites of MW-mass halos \citep[e.g.,][]{GarrisonKimmel2014b, JiangVanDenBosch2015}.
Thus, \textit{our dark-matter-only simulation suffers from the ``too big to fail'' problem.}

We next examine dwarf galaxies in the baryonic simulation.
\fig{mass_function} (left) shows the cumulative number of satellites above a given $\Mstar$.
Blue curves show the baryonic simulation, while black curves show satellites around the MW (dashed) and M31 (dotted).
Latte's satellites span $\Mstar \! = \! 8 \times 10 ^ 4 \! - \! 2 \times 10 ^ 8 \Msun$ and lie entirely between the MW and M31, so \textit{the baryonic simulation does not suffer from ``missing satellites'' at these masses}.

\fig{vel.circ_v_radius} (right) shows the profiles of stellar 3D velocity dispersion, $\sigmavelstar$, for each satellite, demonstrating their flatness.
\fig{mass_function} (right) then shows the cumulative number of satellites above $\sigmavelstar$, as measured at the half-light radius, where it is expected to reflect the total dynamical mass \citep{Walker2009}.
Our high spatial resolution allows us to measure this directly, without uncertainties from extrapolating $\vcirc(r)$ profiles.
\fig{mass_function} compares directly against observed dispersions from \citet{Wolf2010}, converting them to 3D via $\sigma_\textnormal{3D} \! = \! \sqrt{3} \, \sigma_\textnormal{1D}$.
Latte's $\sigmavelstar$ distribution spans $8 \! - \! 35 \kms$ and lies between the MW and M31, so \textit{the baryonic simulation does not suffer from ``too big to fail''}.

For comparison, thin curves in \fig{mass_function} (right) show the distribution of $\Vcircmax$ for dark-matter subhalos in the baryonic (light blue) and dark-matter-only (DMO; orange) simulations.
The baryonic simulation contains $\sim \! 3 \times$ fewer subhalos at fixed $\Vcircmax$.
This significant reduction is driven largely by tidal shocking/stripping from the host's stellar disk (e.g., \citealt{Read2006}; \citealt{Zolotov2012}; Garrison-Kimmel et~al., in prep.).
Furthermore, Latte's (massive) satellites have similar $\sigmavelstar$ and $\Vcircmax$, because FIRE's feedback reduces the dark-matter mass in the core \citep{Chan2015}.

Next, we further demonstrate that Latte's dwarf galaxies have realistic properties.
\fig{property_v_mass} (top) shows $\sigmavelstar$ versus $\Mstar$, for satellite (blue) and isolated (orange) galaxies from Latte (circles) and observations (stars).
All of Latte's galaxies lie within the observed scatter, though Latte's satellites have somewhat larger scatter to low $\sigmavelstar$, likely driven by tidal effects \citep[e.g.,][]{Zolotov2012}, as we will examine in future work.
Overall, $\sigmavelstar$ in \textit{both} satellite and isolated galaxies agrees well with observations across the $\Mstar$ range, primarily because feedback reduces dark-matter densities.
This result is equally important, because isolated low-mass halos in dark-matter-only simulations also suffer from a ``too big to fail'' problem \citep{GarrisonKimmel2014b}.
Thus, \textit{neither satellite nor isolated dwarf galaxies in Latte suffer from a ``too big to fail'' problem}.

The small circles in \fig{property_v_mass} show the 3 isolated and 6 satellite galaxies in the lower-resolution simulation, demonstrating that it resolves galaxies down to $\Mstar \! = \! 4 \times 10 ^ 5 \Msun$.
Furthermore, we find that the $\Mstar \! - \! \Mthm$ relation is nearly identical for (isolated) galaxies in the lower- and higher-resolution simulations above this limit.
However, the overall smaller number of galaxies in the lower-resolution simulation implies that simulations at this (still high) level of resolution, comparable to the MW-mass halos in \citet{Hopkins2014a}, struggle to resolve satellites in a MW-mass halo.
However, the similarity in the $\sigmavelstar \! - \! \Mstar$ relation at both resolutions demonstrates that $\sigmavelstar$ is well resolved in the higher-resolution simulation.

We also examine chemical enrichment via the mass-metallicity relation.
Our simulation generates metals via core-collapse supernovae, Ia supernovae, and stellar winds.
\fig{property_v_mass} (bottom) shows the stellar iron abundance scaled to solar, $\FeH$, for both satellite and isolated galaxies.
Stars show observations in \citet{Kirby2013b}.
As observed, Latte's galaxies have a reasonably tight $\FeH \! - \! \Mstar$ relation.
Three satellites have higher $\FeH$, though they remain close to observed values.
Aside from these three, satellite versus isolated galaxies show \textit{no} systematic offset in $\FeH$, despite systematic differences in star-formation histories.
While Latte's $\FeH$ agree reasonably with observations at $\Mstar \! \gtrsim \! 5 \times 10 ^ 6 \Msun$, they are $\approx \! 0.5 \dex$ low at lower $\Mstar$.
This is a systematic of dwarf galaxies resolved with too few ($\lesssim \! 100$) star particles, possibly from excessively coherent feedback bursts and/or inadequate metallicity sampling in galactic gas (see Hopkins et al., in prep.).
Indeed, galaxies in the lower-resolution simulation (small circles) have even lower $\FeH$, while previous FIRE simulations of isolated dwarf galaxies at higher resolution agreed better with observations \citep{Ma2016}.

Finally, \fig{star_formation_history} shows the cumulative star-formation histories of Latte's satellites, $\Mstar(z)$, computed from their stellar populations at $z \! = \! 0$, along with observed MW satellites from \citet{Weisz2014a}.
Consistent with observations, Latte's satellites show a broad range of histories, and those with higher $\Mstar(z \! = \! 0)$ formed preferentially later.
All satellites at $\Mstar(z \! = \! 0) \! < \! 10 ^ 8 \Msun$ had their star formation quenched (stopped) $3 \! - \! 11 \Gyr$ ago, well after cosmic reionization ($z \! > \! 6$).
However, the most massive satellite remains star-forming to $z \! = \! 0$, broadly consistent with the MW, in which only the most massive satellites (LMC and SMC) remaining star-forming.
Also consistent with the LG, and previous FIRE simulations, almost all of Latte's isolated dwarf galaxies remain star-forming to $z \! \sim \! 0$.
However, 3 at $\Mstar \! \lesssim \! 2 \times 10 ^ 5 \Msun$ do quench by $z \! \sim \! 2$, likely from a strong burst of feedback and/or the ultraviolet background.
In Wetzel et~al., in prep.~we will explore in detail the effects of cosmic reionization, feedback, and environment on these star-formation histories.

\section{Conclusion}

We presented the first results from the Latte Project: an unprecedentedly high-resolution simulation of a MW-mass galaxy within $\Lambda$CDM cosmology, run using \textsc{GIZMO} with the FIRE-2 model for star formation/feedback.
Latte produces a realistic population of satellite and isolated dwarf galaxies, consistent with several observations within the LG: (1) distributions of stellar masses and velocity dispersions (dynamical masses), including their joint relation; (2) the $\Mstar$-stellar metallicity relation; and (3) a diverse range of star-formation histories, including dependence on $\Mstar$.
Critically, Latte's dwarf galaxies do \textit{not} suffer the ``missing satellites'' or ``too big to fail'' problems, down to $\Mstar \! \gtrsim \! 10 ^ 5 \Msun$ and $\sigmavelstar \! \gtrsim \! 8 \kms$.
Because the dark-matter-only simulation suffers from both, we conclude that baryonic physics can account for these observations and thus reconcile dwarf galaxies with standard $\Lambda$CDM cosmology.

This observational agreement arises for primarily two reasons.
First, as demonstrated for isolated dwarf galaxies, FIRE's stellar feedback can generate dark-matter cores \citep{Onorbe2015, Chan2015}, reducing dynamical masses and thus stellar velocity dispersions.
Second, the baryonic simulation contains significantly ($\approx \! 3 \times$) fewer subhalos at fixed $\Vcircmax$ within $d_\host \! < \! 300 \kpc$ than dark-matter-only, because the host's stellar disk destroys subhalos, as we quantify in Garrison-Kimmel et~al., in prep.

We find no significant discrepancies with observed dwarf galaxies, at least at $\Mstar\!\gtrsim\!10^6\Msun$, where Latte resolves star-formation/enrichment histories well.
We will examine additional properties of dwarf galaxies in future works, to further explore both these successes and any potential discrepancies.

\acknowledgments{
We thank Andrew Benson, Mike Boylan-Kolchin, James Bullock, Aflis Deason, Shea Garrison-Kimmel, Marla Geha, Evan Kirby, Robyn Sanderson, Josh Simon, Erik Tollerud, Risa Wechsler for enlightening discussions, Dan Weisz for sharing observations, and Peter Behroozi for sharing \textsc{rockstar}.
We acknowledge support from: Moore Center for Theoretical Cosmology and Physics at Caltech (A.R.W.); Sloan Research Fellowship, NASA ATP grant NNX14AH35G, NSF Collaborative Research grant 1411920 and CAREER grant 1455342 (P.F.H.); Einstein Postdoctoral Fellowship, NASA grant PF4-150147 (J.K.); NSF grants AST-1412836 and AST-1517491, NASA grant NNX15AB22G, and STScI grant HST-AR-14293.001-A (C.-A.F.-G.); NSF grant AST-1412153 and funds from UCSD (D.K.); NASA ATP grant 12-APT12-0183 and Simons Foundation Investigator award (E.Q.).
We used computational resources from the Extreme Science and Engineering Discovery Environment (XSEDE), supported by NSF.
A.R.W. also acknowledges support from lattes.
}


\begin{thebibliography}{}

\bibitem[{{Agertz} \& {Kravtsov}(2015)}]{AgertzKravtsov2015}
{Agertz}, O., \& {Kravtsov}, A.~V. 2015, ArXiv e-prints, arXiv:1509.00853

\bibitem[{{Behroozi} {et~al.}(2013){Behroozi}, {Wechsler}, \&
  {Wu}}]{Behroozi2013a}
{Behroozi}, P.~S., {Wechsler}, R.~H., \& {Wu}, H.-Y. 2013, \apj, 762, 109

\bibitem[{{Boylan-Kolchin} {et~al.}(2011){Boylan-Kolchin}, {Bullock}, \&
  {Kaplinghat}}]{BoylanKolchin2011}
{Boylan-Kolchin}, M., {Bullock}, J.~S., \& {Kaplinghat}, M. 2011, \mnras, 415,
  L40

\bibitem[{{Brooks} \& {Zolotov}(2014)}]{BrooksZolotov2014}
{Brooks}, A.~M., \& {Zolotov}, A. 2014, \apj, 786, 87

\bibitem[{{Busha} {et~al.}(2011){Busha}, {Wechsler}, {Behroozi}, {Gerke},
  {Klypin}, \& {Primack}}]{Busha2011b}
{Busha}, M.~T., {Wechsler}, R.~H., {Behroozi}, P.~S., {et~al.} 2011, \apj, 743,
  117

\bibitem[{{Chan} {et~al.}(2015){Chan}, {Kere{\v s}}, {O{\~n}orbe}, {Hopkins},
  {Muratov}, {Faucher-Gigu{\`e}re}, \& {Quataert}}]{Chan2015}
{Chan}, T.~K., {Kere{\v s}}, D., {O{\~n}orbe}, J., {et~al.} 2015, \mnras, 454,
  2981

\bibitem[{{Dav{\'e}} {et~al.}(2016){Dav{\'e}}, {Thompson}, \&
  {Hopkins}}]{Dave2016}
{Dav{\'e}}, R., {Thompson}, R.~J., \& {Hopkins}, P.~F. 2016, ArXiv e-prints,
  arXiv:1604.01418

\bibitem[{{Di Cintio} {et~al.}(2014){Di Cintio}, {Brook}, {Macci{\`o}},
  {Stinson}, {Knebe}, {Dutton}, \& {Wadsley}}]{DiCintio2014}
{Di Cintio}, A., {Brook}, C.~B., {Macci{\`o}}, A.~V., {et~al.} 2014, \mnras,
  437, 415

\bibitem[{{El-Badry} {et~al.}(2016){El-Badry}, {Wetzel}, {Geha}, {Hopkins},
  {Kere{\v s}}, {Chan}, \& {Faucher-Gigu{\`e}re}}]{ElBadry2016}
{El-Badry}, K., {Wetzel}, A., {Geha}, M., {et~al.} 2016, \apj, 820, 131

\bibitem[{{Faucher-Gigu{\`e}re} {et~al.}(2009){Faucher-Gigu{\`e}re}, {Lidz},
  {Zaldarriaga}, \& {Hernquist}}]{FaucherGiguere2009}
{Faucher-Gigu{\`e}re}, C.-A., {Lidz}, A., {Zaldarriaga}, M., \& {Hernquist}, L.
  2009, \apj, 703, 1416

\bibitem[{{Ferland} {et~al.}(2013){Ferland}, {Porter}, {van Hoof}, {Williams},
  {Abel}, {Lykins}, {Shaw}, {Henney}, \& {Stancil}}]{Ferland2013}
{Ferland}, G.~J., {Porter}, R.~L., {van Hoof}, P.~A.~M., {et~al.} 2013, RMxAA,
  49, 137

\bibitem[{{Flores} \& {Primack}(1994)}]{FloresPrimack1994}
{Flores}, R.~A., \& {Primack}, J.~R. 1994, \apjl, 427, L1

\bibitem[{{Garrison-Kimmel} {et~al.}(2014){Garrison-Kimmel}, {Boylan-Kolchin},
  {Bullock}, \& {Kirby}}]{GarrisonKimmel2014b}
{Garrison-Kimmel}, S., {Boylan-Kolchin}, M., {Bullock}, J.~S., \& {Kirby},
  E.~N. 2014, \mnras, 444, 222

\bibitem[{{Garrison-Kimmel} {et~al.}(2016){Garrison-Kimmel}, {Bullock},
  {Boylan-Kolchin}, \& {Bardwell}}]{GarrisonKimmel2016}
{Garrison-Kimmel}, S., {Bullock}, J.~S., {Boylan-Kolchin}, M., \& {Bardwell},
  E. 2016, ArXiv e-prints, arXiv:1603.04855

\bibitem[{{Hahn} \& {Abel}(2011)}]{HahnAbel2011}
{Hahn}, O., \& {Abel}, T. 2011, \mnras, 415, 2101

\bibitem[{{Hopkins}(2015)}]{Hopkins2015}
{Hopkins}, P.~F. 2015, \mnras, 450, 53

\bibitem[{{Hopkins} {et~al.}(2014){Hopkins}, {Kere{\v s}}, {O{\~n}orbe},
  {Faucher-Gigu{\`e}re}, {Quataert}, {Murray}, \& {Bullock}}]{Hopkins2014a}
{Hopkins}, P.~F., {Kere{\v s}}, D., {O{\~n}orbe}, J., {et~al.} 2014, \mnras,
  445, 581

\bibitem[{{Hopkins} {et~al.}(2013){Hopkins}, {Narayanan}, \&
  {Murray}}]{Hopkins2013c}
{Hopkins}, P.~F., {Narayanan}, D., \& {Murray}, N. 2013, \mnras, 432, 2647

\bibitem[{{Jiang} \& {van den Bosch}(2015)}]{JiangVanDenBosch2015}
{Jiang}, F., \& {van den Bosch}, F.~C. 2015, \mnras, 453, 3575

\bibitem[{{Kirby} {et~al.}(2013){Kirby}, {Cohen}, {Guhathakurta}, {Cheng},
  {Bullock}, \& {Gallazzi}}]{Kirby2013b}
{Kirby}, E.~N., {Cohen}, J.~G., {Guhathakurta}, P., {et~al.} 2013, \apj, 779,
  102

\bibitem[{{Klypin} {et~al.}(1999){Klypin}, {Kravtsov}, {Valenzuela}, \&
  {Prada}}]{Klypin1999b}
{Klypin}, A., {Kravtsov}, A.~V., {Valenzuela}, O., \& {Prada}, F. 1999, \apj,
  522, 82

\bibitem[{{Leitherer} {et~al.}(1999){Leitherer}, {Schaerer}, {Goldader},
  {Delgado}, {Robert}, {Kune}, {de Mello}, {Devost}, \&
  {Heckman}}]{Leitherer1999}
{Leitherer}, C., {Schaerer}, D., {Goldader}, J.~D., {et~al.} 1999, \apjs, 123,
  3

\bibitem[{{Lewis} {et~al.}(2015){Lewis}, {Dolphin}, {Dalcanton}, {Weisz},
  {Williams}, {Bell}, {Seth}, {Simones}, {Skillman}, {Choi}, {Fouesneau},
  {Guhathakurta}, {Johnson}, {Kalirai}, {Leroy}, {Monachesi}, {Rix}, \&
  {Schruba}}]{Lewis2015}
{Lewis}, A.~R., {Dolphin}, A.~E., {Dalcanton}, J.~J., {et~al.} 2015, \apj, 805,
  183

\bibitem[{{Licquia} \& {Newman}(2015)}]{LicquiaNewman2015}
{Licquia}, T.~C., \& {Newman}, J.~A. 2015, \apj, 806, 96

\bibitem[{{Lovell} {et~al.}(2014){Lovell}, {Frenk}, {Eke}, {Jenkins}, {Gao}, \&
  {Theuns}}]{Lovell2014}
{Lovell}, M.~R., {Frenk}, C.~S., {Eke}, V.~R., {et~al.} 2014, \mnras, 439, 300

\bibitem[{{Ma} {et~al.}(2016){Ma}, {Hopkins}, {Faucher-Gigu{\`e}re}, {Zolman},
  {Muratov}, {Kere{\v s}}, \& {Quataert}}]{Ma2016}
{Ma}, X., {Hopkins}, P.~F., {Faucher-Gigu{\`e}re}, C.-A., {et~al.} 2016,
  \mnras, 456, 2140

\bibitem[{{Mashchenko} {et~al.}(2008){Mashchenko}, {Wadsley}, \&
  {Couchman}}]{Mashchenko2008}
{Mashchenko}, S., {Wadsley}, J., \& {Couchman}, H.~M.~P. 2008, Science, 319,
  174

\bibitem[{{McConnachie}(2012)}]{McConnachie2012}
{McConnachie}, A.~W. 2012, \aj, 144, 4

\bibitem[{{Mollitor} {et~al.}(2015){Mollitor}, {Nezri}, \&
  {Teyssier}}]{Mollitor2015}
{Mollitor}, P., {Nezri}, E., \& {Teyssier}, R. 2015, \mnras, 447, 1353

\bibitem[{{Moore}(1994)}]{Moore1994}
{Moore}, B. 1994, \nat, 370, 629

\bibitem[{{Moore} {et~al.}(1999){Moore}, {Ghigna}, {Governato}, {Lake},
  {Quinn}, {Stadel}, \& {Tozzi}}]{Moore1999}
{Moore}, B., {Ghigna}, S., {Governato}, F., {et~al.} 1999, \apjl, 524, L19

\bibitem[{{Muratov} {et~al.}(2015){Muratov}, {Kere{\v s}},
  {Faucher-Gigu{\`e}re}, {Hopkins}, {Quataert}, \& {Murray}}]{Muratov2015}
{Muratov}, A.~L., {Kere{\v s}}, D., {Faucher-Gigu{\`e}re}, C.-A., {et~al.}
  2015, \mnras, 454, 2691

\bibitem[{{O{\~n}orbe} {et~al.}(2015){O{\~n}orbe}, {Boylan-Kolchin}, {Bullock},
  {Hopkins}, {Kere{\v s}}, {Faucher-Gigu{\`e}re}, {Quataert}, \&
  {Murray}}]{Onorbe2015}
{O{\~n}orbe}, J., {Boylan-Kolchin}, M., {Bullock}, J.~S., {et~al.} 2015,
  \mnras, 454, 2092

\bibitem[{{O{\~n}orbe} {et~al.}(2014){O{\~n}orbe}, {Garrison-Kimmel}, {Maller},
  {Bullock}, {Rocha}, \& {Hahn}}]{Onorbe2014}
{O{\~n}orbe}, J., {Garrison-Kimmel}, S., {Maller}, A.~H., {et~al.} 2014,
  \mnras, 437, 1894

\bibitem[{{Oh} {et~al.}(2011){Oh}, {Brook}, {Governato}, {Brinks}, {Mayer}, {de
  Blok}, {Brooks}, \& {Walter}}]{Oh2011}
{Oh}, S.-H., {Brook}, C., {Governato}, F., {et~al.} 2011, \aj, 142, 24

\bibitem[{{Pontzen} \& {Governato}(2012)}]{PontzenGovernato2012}
{Pontzen}, A., \& {Governato}, F. 2012, \mnras, 421, 3464

\bibitem[{{Read} \& {Gilmore}(2005)}]{ReadGilmore2005}
{Read}, J.~I., \& {Gilmore}, G. 2005, \mnras, 356, 107

\bibitem[{{Read} {et~al.}(2006){Read}, {Wilkinson}, {Evans}, {Gilmore}, \&
  {Kleyna}}]{Read2006}
{Read}, J.~I., {Wilkinson}, M.~I., {Evans}, N.~W., {Gilmore}, G., \& {Kleyna},
  J.~T. 2006, \mnras, 367, 387

\bibitem[{{Rocha} {et~al.}(2013){Rocha}, {Peter}, {Bullock}, {Kaplinghat},
  {Garrison-Kimmel}, {O{\~n}orbe}, \& {Moustakas}}]{Rocha2013}
{Rocha}, M., {Peter}, A.~H.~G., {Bullock}, J.~S., {et~al.} 2013, \mnras, 430,
  81

\bibitem[{{Sawala} {et~al.}(2016){Sawala}, {Frenk}, {Fattahi}, {Navarro},
  {Bower}, {Crain}, {Dalla Vecchia}, {Furlong}, {Helly}, {Jenkins}, {Oman},
  {Schaller}, {Schaye}, {Theuns}, {Trayford}, \& {White}}]{Sawala2016}
{Sawala}, T., {Frenk}, C.~S., {Fattahi}, A., {et~al.} 2016, \mnras, 457, 1931

\bibitem[{{Simon} {et~al.}(2005){Simon}, {Bolatto}, {Leroy}, {Blitz}, \&
  {Gates}}]{Simon2005}
{Simon}, J.~D., {Bolatto}, A.~D., {Leroy}, A., {Blitz}, L., \& {Gates}, E.~L.
  2005, \apj, 621, 757

\bibitem[{{Springel}(2005)}]{Springel2005e}
{Springel}, V. 2005, \mnras, 364, 1105

\bibitem[{{Tamm} {et~al.}(2012){Tamm}, {Tempel}, {Tenjes}, {Tihhonova}, \&
  {Tuvikene}}]{Tamm2012}
{Tamm}, A., {Tempel}, E., {Tenjes}, P., {Tihhonova}, O., \& {Tuvikene}, T.
  2012, \aap, 546, A4

\bibitem[{{Tollerud} {et~al.}(2011){Tollerud}, {Boylan-Kolchin}, {Barton},
  {Bullock}, \& {Trinh}}]{Tollerud2011}
{Tollerud}, E.~J., {Boylan-Kolchin}, M., {Barton}, E.~J., {Bullock}, J.~S., \&
  {Trinh}, C.~Q. 2011, \apj, 738, 102

\bibitem[{{Walker} {et~al.}(2009){Walker}, {Mateo}, {Olszewski},
  {Pe{\~n}arrubia}, {Wyn Evans}, \& {Gilmore}}]{Walker2009}
{Walker}, M.~G., {Mateo}, M., {Olszewski}, E.~W., {et~al.} 2009, \apj, 704,
  1274

\bibitem[{{Weisz} {et~al.}(2014){Weisz}, {Dolphin}, {Skillman}, {Holtzman},
  {Gilbert}, {Dalcanton}, \& {Williams}}]{Weisz2014a}
{Weisz}, D.~R., {Dolphin}, A.~E., {Skillman}, E.~D., {et~al.} 2014, \apj, 789,
  147

\bibitem[{{Wetzel} {et~al.}(2015){Wetzel}, {Tollerud}, \&
  {Weisz}}]{Wetzel2015b}
{Wetzel}, A.~R., {Tollerud}, E.~J., \& {Weisz}, D.~R. 2015, \apjl, 808, L27

\bibitem[{{Wheeler} {et~al.}(2015){Wheeler}, {Pace}, {Bullock},
  {Boylan-Kolchin}, {Onorbe}, {Fitts}, {Hopkins}, \& {Keres}}]{Wheeler2015b}
{Wheeler}, C., {Pace}, A.~B., {Bullock}, J.~S., {et~al.} 2015, ArXiv e-prints,
  arXiv:1511.01095

\bibitem[{{Wolf} {et~al.}(2010){Wolf}, {Martinez}, {Bullock}, {Kaplinghat},
  {Geha}, {Mu{\~n}oz}, {Simon}, \& {Avedo}}]{Wolf2010}
{Wolf}, J., {Martinez}, G.~D., {Bullock}, J.~S., {et~al.} 2010, \mnras, 406,
  1220

\bibitem[{{Zolotov} {et~al.}(2012){Zolotov}, {Brooks}, {Willman}, {Governato},
  {Pontzen}, {Christensen}, {Dekel}, {Quinn}, {Shen}, \&
  {Wadsley}}]{Zolotov2012}
{Zolotov}, A., {Brooks}, A.~M., {Willman}, B., {et~al.} 2012, \apj, 761, 71

\end{thebibliography}

\end{document}